\documentclass[11pt]{article}       
\usepackage{amsmath}
\usepackage{amsfonts}
\usepackage{amssymb}
\usepackage{lineno}
\usepackage{graphicx}
\usepackage{color}
\usepackage{natbib}
\usepackage{apalike}
\usepackage{bbm}
\usepackage{multirow}
\usepackage{booktabs}
\usepackage{xcolor}
\usepackage{setspace}
\usepackage{caption}
\usepackage{float}
\usepackage[font=footnotesize, labelfont=bf]{caption}
\usepackage{eurosym}
\usepackage{placeins}
\usepackage{bigints}
\newtheorem{teo1}{Algorithm}
\newtheorem{teo2}[teo1]{Algorithm}

\newtheorem{teo5}{Condition}
\newtheorem{teo6}{Proposition}
\newtheorem{teo7}[teo6]{Proposition}

\begin{document}

\title{Unsupervised Mixture Estimation via Approximate Maximum Likelihood based on the Cramér - von Mises distance}

\author{Marco Bee \\
	{\it\small Department of Economics and Management, University of Trento - Italy} 
}

\title{Unsupervised Mixture Estimation via Approximate Maximum Likelihood based on the Cramér - von Mises distance}

\author{Marco Bee \\
	{\it\small Department of Economics and Management, University of Trento}} 

\date{ }

\maketitle

\begin{abstract}
Mixture distributions with dynamic weights are an efficient way of modeling loss data characterized by heavy tails. However, maximum likelihood estimation of this family of models is difficult, mostly because of the need to evaluate numerically an intractable normalizing constant. In such a setup, simulation-based estimation methods are an appealing alternative. The approximate maximum likelihood estimation (AMLE) approach is employed. It is a general method that can be applied to mixtures with any component densities, as long as simulation is feasible. The focus is on the dynamic lognormal-generalized Pareto distribution, and the Cramér - von Mises distance is used to measure the discrepancy between observed and simulated samples. After deriving the theoretical properties of the estimators, a hybrid procedure is developed, where standard maximum likelihood is first employed to determine the bounds of the uniform priors required as input for AMLE. Simulation experiments and two real-data applications suggest that this approach yields a major improvement with respect to standard maximum likelihood estimation.
\end{abstract}

\noindent{\textbf{Keywords}: Dynamic Mixtures; Simulation; Classification; Tail events; Statistical distance.

\bigskip

\section{Introduction}
\label{sec:intro}

Estimating the right tail of a non-negative probability distribution is very important in many fields such as hydrology, economics and finance. At the same time, the problem may be difficult from the modeling point of view, since the data-generating processes of the tail and the body are often different, so that no single stochastic model guarantees a precise description of the entire distribution. On top of that, statistical inference for the tail is challenging, because samples of extreme observations are typically small.

A fundamental set of tools for estimating the tail is known as Extreme Value Theory (EVT), which focuses on the distribution of the largest observations, or of the excesses above a high threshold. 
However, EVT only fits the tail; if interest is in the whole distribution, it is advisable to resort to a spliced distribution combining two different models for the body and the tail. A possible drawback is the need of continuity and differentiability constraints that reduce the number of free parameters \citep{scoll07}.

An even more flexible alternative is the dynamic model developed by \cite{fri02}. This approach employs a non-negative distribution with a size distribution supported on $[0,\infty)$ for the body and the Generalized Pareto distribution (GPD) for the tail. Unlike classical finite mixtures, the density $f(x)$ of the resulting random variable $X$ is based on a dynamic weight, which is a monotonically increasing function of $x$. In this way, the GPD is given more and more weight as we move farther into the right tail, and the tail behavior is governed by the GPD. If the weight function is continuous, the density $f$ is continuous as well. Finally, this is an unsupervised method, because no threshold needs to be chosen or estimated.

Despite its interesting properties, to the best of our knowledge this model has not been used in practice. The likely reason is that maximum likelihood estimation (MLE) is rather complex. On one hand, the mixing weight and the component densities share all parameters, so that, as pointed out by \cite{fri02}, the use of the EM algorithm is precluded. On the other hand, direct maximization of the likelihood is feasible, but the normalizing constant is given by an integral that cannot be computed explicitly, and whose numerical approximation is not trivial. 

Given the difficulties related to MLE, simulation-based methods are an appealing alternative that avoids the evaluation of the normalizing constant, since there is no optimization of the likelihood function. Hence, the possible error caused by an inaccurate approximation of the normalizing constant is eliminated, at the price of the introduction of simulation error.

In this paper we explore the second strategy, and use the Approximate Maximum Likelihood Estimation (AMLE) method proposed by \cite{rub13} and applied to similar setups by \cite{bee15}, \cite{bee17} and \cite{beetaf22}. The approach is essentially a frequentist version of the family of Approximate Bayesian Computation (ABC) techniques, aimed at approximate estimation of the posterior distributions of model parameters; see, e.g., \cite{sun13}, \cite{bea19}, \cite{dro22}. In the early stages of ABC, the crucial issue was the choice of the summary statistics used to assess the proximity of true and simulated data. This issue is shared by AMLE and is not easy to address in general, except for uncommon setups where sufficient statistics are available.

To overcome this problem, in recent years several papers have developed ABC versions that compare empirical distributions of observed and simulated data, thus bypassing the need of choosing summary statistics. This way of proceeding was first proposed by \cite{park16}, who employed a non-parametric approach; a more comprehensive treatment was given by \cite{ber19}. See \cite{dro22} and the references therein for a thorough review.
In the following, we will term these ways of proceeding \emph{full-data approaches}, to emphasize the use of all the data, with no dimensionality reduction transformation. Among the various proposals in the literature, we base our implementation on the Cramér-von Mises (CvM) distance, which is a simple and effective measure.

AMLE and MLE are compared via simulation: our outcomes suggest that AMLE has smaller root-mean-squared-error (RMSE) in all setups considered in the experiments. Two empirical analyses confirm the suitability of the approach for modeling and estimating skewed and fat-tailed distributions. Furthermore, the estimated weight allows one to identify the approximate number of observations generated by the heavy-tailed component.

On the theoretical side, full-data AMLE yields a likelihood approximation that converges pointwise to the posterior distribution. Under slightly stronger conditions, the mode of the approximation converges pointwise to the mode of the likelihood. 

The rest of the paper is organized as follows. In Section \ref{sec:model} we describe the dynamic mixture distribution. Section \ref{sec:est} contains a detailed account of the AMLE versions with and without summary statistics. In sections \ref{sec:MC} and \ref{sec:emp} we report the outcomes of the simulation experiments and of the empirical analysis, respectively. Finally, Section \ref{sec:concl} concludes the paper and outlines open problems and possible further developments.

\section{The dynamic mixture model}
\label{sec:model}

The density of a dynamic mixture model is given by
\begin{equation}
	f(x;\boldsymbol{\theta})=\frac{(1-p(x;\boldsymbol{\theta}_0))f_1(x;\boldsymbol{\theta}_1)+p(x;\boldsymbol{\theta}_0)f_2(x;\boldsymbol{\theta}_2)}{Z},\quad x\in\mathbb{R}^+,
	\label{eq:dens0}
\end{equation}
where $\boldsymbol{\theta}=(\boldsymbol{\theta}_0,\boldsymbol{\theta}_1,\boldsymbol{\theta}_2)$, $Z$ is a normalizing constant and $\boldsymbol{\theta}_i$, $i=0,1,2$, are the parameter vectors of the weights $p(x;\boldsymbol{\theta}_0)$, the body $f_1(x;\boldsymbol{\theta}_1)$ and the tail $f_2(x;\boldsymbol{\theta}_2)$, respectively. For the weight function it is convenient to employ the cumulative distribution function (cdf) of some continuous random variable, whereas $f_1$ and $f_2$ are continuous densities with positive support. Analogously to \cite{fri02}, here $p(x;\boldsymbol{\theta}_0)$ is taken to be equal to the Cauchy cdf:
$$
p(x;\mu_c,\tau)=\frac{1}{2}+\frac{1}{\pi}\arctan\left(\frac{x-\mu_c}{\tau}\right),
$$
so that $\boldsymbol{\theta}_0=(\mu_c,\tau)$, where $\mu_c\in\mathbb{R}$ and $\tau\in\mathbb{R}^+$ are the location and scale parameter, respectively.

As for $f_1$ and $f_2$, we use the lognormal and the zero-mean generalized Pareto densities, respectively. The use of the latter for the tail is related to its importance in EVT, since it is the asymptotic distribution of the excesses (see, e.g., \citealp{emb97}). As of the lognormal, it is chosen for two reasons. First, its flexibility can accommodate various possible shapes. Second, in economics there is a long-standing debate about the proper data-generating process of various relevant variables, with theoretical reasons suggesting a lognormal distribution, possibly with a Pareto-type tail. See, e.g., \cite{dac19} and the references therein for city size, \cite{axte01}, \cite{digi11}, \cite{tan15}, \cite{beers17} and \cite{kon21} for firm size.

Accordingly, in the lognormal-GPD case the density (\ref{eq:dens0}) is given by
\begin{equation}
f(x;\boldsymbol{\theta})=\frac{(1-p(x;\mu_c,\tau))f_1(x;\mu,\sigma^2)+p(x;\mu_c,\tau)f_2(x;\beta,\xi)}{Z},
\label{eq:dens}
\end{equation}
where $\mu_c,\mu,\xi\in\mathbb{R}$, $\tau,\sigma^2,\beta\in\mathbb{R}^+$, $\boldsymbol{\theta}=(\mu_c,\tau,\mu,\sigma^2,\beta,\xi)'$, $\boldsymbol{\theta}_1=(\mu,\sigma^2)$, $\boldsymbol{\theta}_2=(\beta,\xi)$,  $f_1(x;\mu,\sigma^2)$ is the lognormal density with parameters $\mu$ and $\sigma^2$, $f_2(x;\beta,\xi)$ is the GPD pdf centered at 0 with scale and shape parameters equal to $\beta$ and $\xi$, respectively. Finally, the normalizing constant $Z$ is equal to
$$
Z=Z(\boldsymbol{\theta})=1+\frac{1}{\pi}I,
$$
where
\begin{equation}
\label{eq:int}
I=\bigintsss_0^\infty\left[\frac{1}{\beta}\left(1+\frac{\xi x}{\beta}\right)^{-1/\xi-1}-\frac{1}{\sqrt{2\pi}\sigma x}e^{-\frac{1}{2}\left(\frac{\log{x}-\mu}{\sigma}\right)^2}\right]\arctan\left(\frac{x-\mu_c}{\tau}\right)dx.
\end{equation}
See \cite{fri02} for details. 

From a modeling point of view, in the literature there are at least two possible alternatives to (\ref{eq:dens}), both based on the Pareto instead of the GPD.
\cite{scoll07} proposes a lognormal-Pareto mixture constrained to have a continuous and differentiable density. The constraints reduce the number of free parameters by two; in particular, the mixing weight becomes a function of the lognormal variance and of the parameters of the Pareto distribution (\citealp{scoll07,bee15a}). Another possibility is the use of (\ref{eq:dens}) with a mixing weight $p(x)$ equal to the Heavyside function. In this setup, the density is no longer continuous; see \cite{bee22a} for the analysis of this case.

\section{Estimation}
\label{sec:est}

In the rest of the paper, we focus on the lognormal-GPD dynamic mixture, but other models corresponding to different component densities can be estimated via the same method. The only requirement for implementing the proposed AMLE approach is indeed the ability to simulate from the two distributions of the mixture.

\subsection{Maximum likelihood}
\label{sec:ML}

As usual, maximum likelihood estimators can be found by maximizing the log-likeli\-hood function obtained by taking the natural logarithm of (\ref{eq:dens}):
\begin{equation}
l(\boldsymbol{\theta};\boldsymbol{x})=\sum_{i=1}^n\log\left\{ \frac{(1-p(x_i;\mu_c,\tau))f_1(x_i;\mu,\sigma^2)+p(x_i;\mu_c,\tau)f_2(x_i;\beta,\xi)}{Z}\right\}.
\label{eq:llik}
\end{equation}
Since the normalizing constant $Z$ depends on all parameters, the evaluation of (\ref{eq:int}) is crucial for MLE. 
To achieve a sufficiently high level of precision in the approximation, \cite{fri02} suggest to split the integral on $[0,\infty)$ in a sum of integrals on finite intervals, such as $[0, 1]$, $[1, 2]$, and so on.

Numerical maximization of (\ref{eq:llik}) is implemented in the \texttt{OpVaR} \texttt{R} package, where the \texttt{dmixing} function 
carries out a numerical approximation of the whole improper integral from 0 to $\infty$. On the contrary, we have tackled the evaluation via the \texttt{quadinf} function of the \texttt{pracma R} package on non-overlapping intervals $[n-1,n]$, $n\in\mathbbm{N}$. The stopping criterion is $I_{n-1,n}<\epsilon_I$ with $\epsilon_I=10^{-4}$, where $I_{n-1,n}$ is the value of the integral on $[n-1,n]$. 
An unreported numerical comparison between the \texttt{OpVaR} and our procedure produced considerably better outcomes, especially in terms of variance, for the latter, which will always be used in the following.

The tolerance $\epsilon_I$ that defines the stopping criterion affects both the accuracy of the numerical evaluation of the normalizing constant and the computational burden of MLE. The actual value  $\epsilon_I=10^{-4}$ employed in the following has been determined via simulation (see Sect. \ref{sec:MC_des}).

\subsection{Approximate Maximum Likelihood: some background}

The Approximate Maximum Likelihood method is a simulation-based estimation procedure \citep{rub13}. We give a general description in this section, and then detail the full-data algorithm (Section \ref{sec:without}) and the approach based on summary statistics (Section \ref{sec:with}). Finally, Section \ref{sec:last} describes the last stage of the procedure, which is common to both cases.

Given a sample $\boldsymbol{x}=(x_1,\dots,x_n)'\in\mathbb{R}^{q\times n}$ from a distribution with density $f(\boldsymbol{x};\boldsymbol{\theta})$, let $L(\boldsymbol{\theta};\boldsymbol{x})$ be the likelihood function, where $\boldsymbol{\theta}\in\boldsymbol{\Theta}\subset\mathbbm{R}^p$ is a vector of parameters. We formally introduce the method by assuming a Bayesian setup with a prior distribution $\pi(\boldsymbol{\theta})$. This is done only for mathematical convenience: as discussed below, by using a uniform prior, in the following we carry out a likelihood analysis. The posterior $\pi(\boldsymbol{\theta}|\boldsymbol{x})$ is given by
\begin{equation}
	\label{eq:post}
	\pi(\boldsymbol{\theta}|\boldsymbol{x})=\frac{f(\boldsymbol{x}|\boldsymbol{\theta})\pi(\boldsymbol{\theta})}{\int_{\boldsymbol{\Theta}}f(\boldsymbol{x}|\boldsymbol{t})\pi(\boldsymbol{t})d\boldsymbol{t}}.
\end{equation}
Let's now introduce an approximation of the likelihood defined as follows:
\begin{equation}
	\label{eq:app_lik}
	\hat{f}_\epsilon(\boldsymbol{x}|\boldsymbol{\theta})=\int_{\mathbbm{R}^n}K_\epsilon(\boldsymbol{x}|\boldsymbol{z})f(\boldsymbol{z}|\boldsymbol{\theta})d\boldsymbol{z},
\end{equation}
where $K_\epsilon(\boldsymbol{x}|\boldsymbol{z})$ is a normalized Markov kernel (\citealp[Sect. 6.2]{cas04}; \citealp{hey82}; for the specific form used here, see (\ref{eq:ker}) below) depending on a scale parameter $\epsilon$. Now we plug (\ref{eq:app_lik}) into (\ref{eq:post}) to obtain the following approximation of the posterior:
$$
\hat{\pi}_\epsilon(\boldsymbol{\theta}|\boldsymbol{x})=\frac{\hat{f}_\epsilon(\boldsymbol{x}|\boldsymbol{\theta})\pi(\boldsymbol{\theta})}{\int_{\boldsymbol{\Theta}}\hat{f}_\epsilon(\boldsymbol{x}|\boldsymbol{t})\pi(\boldsymbol{t})d\boldsymbol{t}}.
$$
If the prior $\pi(\boldsymbol{\theta})$ is uniform, the maximization of the likelihood is equivalent to the maximization of the posterior. 

Now we exploit the quantities defined above to give a pseudo-code of the algorithm.

\begin{teo1} \ (AMLE) \\
	\begin{enumerate}
		\item Obtain a sample $\boldsymbol{\theta}_{\epsilon}^*=(\boldsymbol{\theta}_{\epsilon,1}^*,\dots,\boldsymbol{\theta}_{\epsilon,\ell}^*)'$ from the approximate posterior $\hat{\pi}_\epsilon(\boldsymbol{\theta}|\boldsymbol{x})$; $\ell$ is commonly called ABC sample size;
		\item Use this sample to construct a non-parametric estimator $\hat{\phi}$ of the density $\hat{\pi}_\epsilon(\boldsymbol{\theta}|\boldsymbol{x})$;
		\item Compute the maximum of $\hat{\phi}$, $\tilde{\boldsymbol{\theta}}_{\ell,\epsilon}$. This is an approximation of the MLE $\boldsymbol{\hat{\theta}}$.
	\end{enumerate}
	\label{alg:AMLE}
\end{teo1}

Algorithmically, AMLE is slightly different in the full-data approach and in the setup based on summary statistics. In particular, Step 1 is usually based on the ABC rejection algorithm (\citealp{bea10}), whose implementation is not the same in the two frameworks. Accordingly, we provide the details of the two cases in sections \ref{sec:without} and \ref{sec:with}.

\subsubsection{Stage 1a: AMLE without summary statistics}
\label{sec:without}

In a full-data approach, the ABC sample mentioned at Step 1 of Algorithm \ref{alg:AMLE} is obtained as follows.

\begin{teo2} \ (ABC rejection algorithm) \\
	\begin{enumerate}
		\item Simulate $\boldsymbol{\theta}^*$ from the uniform prior $\pi(\cdot)$;
		\item Generate $\boldsymbol{z}=(z_1,\dots,z_n)'$ from $f(\cdot|\boldsymbol{\theta}^*)$;
		\item Accept $\boldsymbol{\theta}^*$ with probability $\propto K_\epsilon(\boldsymbol{x}|\boldsymbol{z})$, otherwise return to Step 1.
	\end{enumerate}
	\label{alg:ABC1}
\end{teo2}

Let $P^{(n)}_{\boldsymbol{\theta}}\in\mathcal{P}$ denote the distribution of $\boldsymbol{z}|\boldsymbol{\theta}$ and $P_0^{(n)}\in\mathcal{P}$ the distribution of $\boldsymbol{x}$, where $\mathcal{P}\stackrel{\textnormal{def}}{=}\{P_{\boldsymbol{\theta}}^{(n)},\ \boldsymbol{\theta}\in\boldsymbol{\Theta}\subset\mathbb{R}^p\}$ is the class of probability measures that is assumed to have generated the data. Furthermore, let $\rho : \mathcal{P} \times\mathcal{P} \to \mathbb{R}^+$ be a statistical distance on $\mathcal{P}$.
The kernel $K_\epsilon$ is defined on the space of the true and simulated data:
\begin{equation}
	\label{eq:ker}
	K_\epsilon(\boldsymbol{x}|\boldsymbol{z})\propto
	\begin{cases}
		1 & \textnormal{if}\ \rho(P_0^{n},P_{\boldsymbol{\theta}}^{n})<\epsilon, \\
		0 & \textnormal{otherwise}.
	\end{cases}
\end{equation}
Since $P_0^{n}$ is unknown and $P_{\boldsymbol{\theta}}^{n}$ is intractable, (\ref{eq:ker}) cannot be computed. The common workaround \citep{dro22} replaces the two distributions with the corresponding empirical counterparts, so that (\ref{eq:ker}) becomes:
$$
	K_\epsilon(\boldsymbol{x}|\boldsymbol{z})\propto
\begin{cases}
	1 & \textnormal{if}\ \rho (\hat{F}(\boldsymbol{x}),\hat{F}_{\boldsymbol{\theta}}(\boldsymbol{z}))<\epsilon, \\
	0 & \textnormal{otherwise},
\end{cases}
$$
where $\hat{F}(\cdot)$ and $\hat{F}_{\boldsymbol{\theta}}(\cdot)$ are the empirical cdfs based on the observed data $\boldsymbol{x}$ and the simulated data $\boldsymbol{z}$, respectively.
As of $\rho$, there are various possibilities; in the following, we take it to be the Cramér-von Mises distance, given by
\begin{equation}
\hat{\mathcal{C}}(\hat{F},\hat{F}_{\boldsymbol{\theta}})=\int_{\mathbb{R}^q}[\hat{F}(t)-\hat{F}_{\boldsymbol{\theta}}(t)]^2d\hat{H}(t),
\label{eq:CvM}
\end{equation}
where $\hat{H}(t)=(\hat{F}(t)+\hat{F}_{\boldsymbol{\theta}}(t))/2$. Note that (\ref{eq:CvM}) can be easily computed in terms of the ranks of the observed and simulated samples. Moreover, it is robust to fat tails and outliers, which is a very desirable property in our application and is therefore the main reason for preferring it; see \citet[p. 7]{dro22} for details and comparisons to other distances.

The full-data approach may give the false impression that ABC (of which AMLE is a reinterpretation) is a method that compares statistical distances between simulated and empirical distributions. However, the main goal of ABC is the approximation of properties of the posterior distribution of parameters. Historically, it has been first employed in population genetics to estimate parameters in setups where the likelihoods of the underlying probability models were intractable. Moreover, such early applications typically used summary statistics. See, e.g., \cite{bea02} or \cite{tav18} for details.

\subsubsection{Stage 1b: AMLE with summary statistics}
\label{sec:with}

In the version of AMLE based on summary statistics, steps 1 and 2 of Algorithm \ref{alg:ABC1} remain unchanged, whereas Step 3 is modified as follows.

\begin{enumerate}
	\item[3.] Use $\boldsymbol{x}$ to compute an $m$-dimensional summary statistics $\boldsymbol{\eta}(\boldsymbol{x})$; accept $\boldsymbol{\theta}^*$ with probability $\propto K^\rho_\epsilon(\boldsymbol{\eta}(\boldsymbol{x})|\boldsymbol{\eta}(\boldsymbol{z}))$, otherwise return to Step 1.
\end{enumerate}
The kernel is now defined on the space of the summary statistics:
$$
	K^\rho_\epsilon(\boldsymbol{\eta}(\boldsymbol{x})|\boldsymbol{\eta}(\boldsymbol{z}))\propto
	\begin{cases}
		1 & \rho(\boldsymbol{\eta}(\boldsymbol{x}),\boldsymbol{\eta}(\boldsymbol{z}))<\epsilon, \\
		0 & \text{otherwise}.
	\end{cases}
$$
Here $\rho:\mathbb{R}^m\times \mathbb{R}^m \to\mathbb{R}^+$ is a metric; usually, but not necessarily, $\rho$ is the Euclidean distance.

\cite{rub13} show that the replacement of the observed sample with a (vector of) summary statistics implies no loss of information if and only if $\boldsymbol{\eta}$ is a jointly sufficient statistic for the unknown parameters of the model: in this case, 
conditioning upon the sufficient statistics is the same as conditioning upon the sample.

\subsection{Stage 2: computing the estimator}
\label{sec:last}

The two approaches outlined in sections \ref{sec:without} and \ref{sec:with} yield a sample $\boldsymbol{\theta}^*_\epsilon$ of size $\ell$. The AMLE approach now exploits this sample to compute the estimate in the same way, regardless of the method used for obtaining  it. Step 2 of Algorithm \ref{alg:AMLE} requires to find the nonparametric estimator $\hat{\phi}$ and its maximum. For this task, we consider the following four techniques (\citealp{bee17}):
\begin{itemize}
\item[(i)] the sample mean vector (``M'');
\item[(ii)] the vector of the maxima of the univariate estimated kernel densities (``UK'');
\item[(iii)] the maximum of the multivariate kernel density (``MK'');
\item[(iv)] the maximum of the product of the univariate kernel densities, estimated using the marginal data (``PUK'');
\end{itemize}
In general, ``M'' is easy to compute but is only appropriate when the distribution of the simulated values is approximately symmetric. ``MK'' would in principle be the best approach, but it requires a large sample size, especially when the parameter space is high-dimensional. Hence, in practice, one should rather employ ``UK'' or ``PUK''. While it is difficult to give a criterion for choosing between them, it is often the case that they yield similar estimates (see, e.g., \citealp{bee22}).

Finally, the maximum of $\hat{\phi}$ computed via any of the methods (i)-(iv), denoted by $\tilde{\boldsymbol{\theta}}_{l,\epsilon}\stackrel{\textnormal{def}}{=}\max_{\boldsymbol{\theta}} \hat{\phi}$, is the AMLE estimator, i.e. the AMLE approximation of the MLE $\boldsymbol{\hat{\theta}}$.

The first step of Algorithm \ref{alg:ABC1} is based on uniform priors whose support must contain the true parameter value. Hence, it cannot be too narrow, or it may not contain the true parameter value. However, if it is too wide, the computational burden of the algorithm becomes unnecessarily large. In the present setup, since we can approximate and maximize numerically the likelihood, we exploit the MLEs to find the proper support of the uniform priors as follows.
\begin{itemize}
\item Compute the MLEs and their standard errors by means of non-parametric bootstrap;
\item Discard possible outliers in each bootstrap distribution;
\item For the lognormal and GPD parameters, the support of the uniform prior is set equal to the 99\% confidence interval of the bootstrap distribution after discarding the outliers;
\item For the Cauchy parameters, the support is given by the range of the bootstrap distribution after discarding the outliers.
\end{itemize}
The reason why the entire range is used for the Cauchy parameters is that, according to our Monte Carlo experiments, MLEs of these two parameters are less precise.

In the procedure outlined above, outliers are identified via the classical box-plot approach \citep{tuk77}. Since the histograms of the bootstrap distributions of the MLEs of $\mu_c$ and of $\tau$ (not shown here to save space) are skewed, we also resorted to the adjusted box-plot proposed by \cite{hub08}. With respect to the classical box-plot, the results only change for the MLEs of the Cauchy distribution parameters: in this case, the supports of the uniform priors are wider, since in presence of skewness the adjusted box-plot finds fewer outliers than the classical one \citep{hub08}. However, in our simulation experiments (see Section \ref{sec:MC}), the supports yielded by the classical box-plot are wide enough (i.e., they contain the true parameter value with a rather large margin), so that wider supports only increase the computational burden. Nevertheless, it may not be generally true that the classical box-plot is preferable, hence it is worth considering the use of the adjusted box-plot, especially when the distributions of the MLEs are very skewed.

\subsection{Asymptotic properties}

The limiting theory for the setup where AMLE is based on a summary statistic $\boldsymbol{\eta}$ is covered in \cite{rub13}: when $\boldsymbol{\eta}$ is non-sufficient, the AMLE approximation converges pointwise, under regularity conditions, to the posterior distribution. See \citet[Proposition 2]{rub13} for details.

In the full-data approach, stronger results hold true. \cite{rub13} prove that, when sufficient statistics are available, under an additional assumption, the mode of the approximation $\hat{\pi}_\epsilon(\boldsymbol{\theta}|\boldsymbol{x})$ converges pointwise to the mode of the likelihood $\pi(\boldsymbol{\theta}|\boldsymbol{x})$. Noting that the original data or its
empirical distribution constitute a sufficient statistic, the present setup turns out to be a special case of the framework where sufficient statistics exist. For the sake of completeness, we state explicitly the main theorem and a corollary; the proofs are omitted, since they are almost identical to those in \cite{rub13}. Before stating the results, we need to introduce the following condition.

\begin{teo5}
(Concentration Condition) A family of symmetric Markov kernels with densities $K_\epsilon$ indexed by $\epsilon > 0$ satisfies the concentration condition if the member densities become increasingly concentrated as $\epsilon$ decreases:
$$
\int_{\mathcal{B}_\epsilon(\boldsymbol{x})}K_\epsilon(\boldsymbol{x}|\boldsymbol{y})d\boldsymbol{y}=\int_{\mathcal{B}_\epsilon(\boldsymbol{x})}K_\epsilon(\boldsymbol{y}|\boldsymbol{x})d\boldsymbol{x}=1,\ \forall \epsilon>0,
$$
where $\mathcal{B}_\epsilon(\boldsymbol{x}) := {\boldsymbol{z} : |\boldsymbol{z} - \boldsymbol{x}| \le \epsilon}$.
\end{teo5}	

Proposition \ref{thm:co1} below shows that the AMLE approximation converges pointwise to the posterior distribution.
\begin{teo6}
\label{thm:co1}
Let $\boldsymbol{x}=(\boldsymbol{x}_1,\dots,\boldsymbol{x}_n)'\in\mathbb{R}^{q\times n}$ be a sample from $f(\cdot|\boldsymbol{\theta})$, $\boldsymbol{\theta}\in\boldsymbol{\Theta}\subset\mathbb{R}^d$, and let $\rho:\mathbb{R}^{q\times n}\times \mathbb{R}^{q\times n}\to\mathbb{R}$ be the CvM distance (\ref{eq:CvM}). Suppose further that $f(\cdot|\boldsymbol{\theta})$ is $\rho$-continuous $\forall\boldsymbol{\theta}\in\boldsymbol{D}$, where $\boldsymbol{D}\subset\mathbb{R}^d$ is compact. Assume also that
$$
\sup_{(\boldsymbol{t},\boldsymbol{\theta})\in\mathcal{B}_\epsilon\times \boldsymbol{D}}f(\cdot|\boldsymbol{\theta})<\infty,
$$
and let 
\begin{equation}
\label{eq:kern}
K_\epsilon(\boldsymbol{x}|\boldsymbol{y})=
\begin{cases}
1 & \textnormal{if}\ \rho(\boldsymbol{x},\boldsymbol{y})<\epsilon,\\
0 & \textnormal{otherwise}.
\end{cases}
\end{equation}
Then, $\forall\boldsymbol{\theta}\in\boldsymbol{D}$ and the kernel (\ref{eq:kern}),
$$
\lim_{\epsilon\to 0}\hat{\pi}_\epsilon(\boldsymbol{\theta}|\boldsymbol{x})=\pi(\boldsymbol{\theta}|\boldsymbol{x}).
$$
\end{teo6}

Under the additional condition of equicontinuity of $\hat{\pi}_\epsilon(\cdot|\boldsymbol{x})$, Proposition 1 can be used to show that the mode of the approximation $\hat{\pi}_\epsilon(\cdot|\boldsymbol{x})$ converges to the mode of the likelihood $\pi(\cdot|\boldsymbol{x})$.

\begin{teo7}
Let $\tilde{\boldsymbol{\theta}}_\epsilon$ be the unique maximum of $\hat{\pi}_\epsilon(\cdot|\boldsymbol{x})$, $\forall\epsilon>0$, and assume that $\pi(\cdot|\boldsymbol{x})$ has a unique maximiser $\tilde{\boldsymbol{\theta}}$. Under the conditions in Proposition 1, and if $\hat{\pi}_\epsilon(\cdot|\boldsymbol{x})$ is equicontinuous on $D$, then
$$
\lim_{\epsilon\to 0}\hat{\pi}_\epsilon(\tilde{\boldsymbol{\theta}}_\epsilon|\boldsymbol{x}) = \pi(\tilde{\boldsymbol{\theta}}|\boldsymbol{x}).
$$
\end{teo7}

\section{Monte Carlo experiments}
\label{sec:MC}


\subsection{Simulation design and numerical details}
\label{sec:MC_des}

Four experiments are run: we draw samples of size $n\in\{100,500\}$ from the dynamic mixture with parameters $\mu_c=1$, $\tau=2$, $\mu = 0$, $\sigma = 0.5$, $\xi \in\{0.25,0.5\}$, $\beta = 3.5$. The densities of the two distributions corresponding to $\xi=0.25$ and $\xi=0.5$ are displayed in Fig. \ref{fig:mix1}.

\begin{figure}[!h]
\begin{center}
\includegraphics[width=\textwidth]{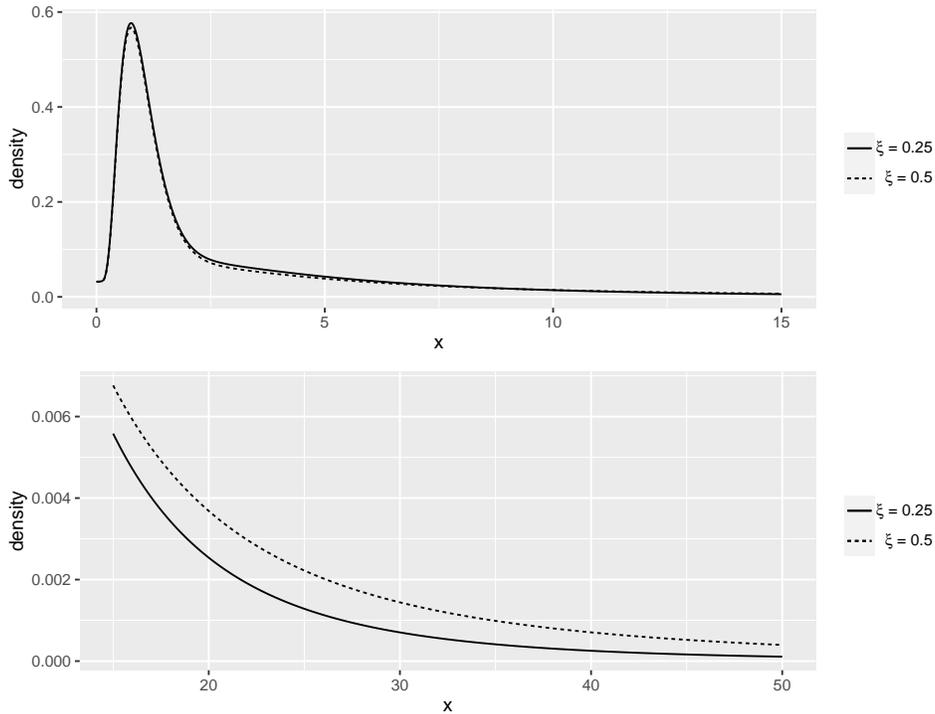}
\caption{Densities of the two simulated distributions. The bodies ($x<15$) are represented in the top panel, the tails ($x\ge 15$) in the bottom panels.}
\label{fig:mix1}
\end{center}
\end{figure}

MLEs result from the maximization of the log-likelihood function by means of the \texttt{optim} \texttt{R} command, where the normalizing constant is approximated as described in Section \ref{sec:ML}. Starting values for $\mu$ and $\sigma$ are the lognormal MLEs computed with the observations below the median. Similarly, initial values for $\xi$ and $\tau$ are the GPD MLEs obtained with the observations above the median. As for $\mu_c$ and $\tau$, finding a reasonable initialization is more complicated. We use $\tau^0=\log(\textnormal{sd}(\boldsymbol{x})/2)$, where sd is the sample standard deviation, and $\mu_c^0=q_{0.25}(\boldsymbol{x})$, where $q_{0.25}$ is the first quartile of the data. The former is identical to the initialization employed in the \texttt{OpVaR} package, whereas the latter is different: the package uses the third quartile, but according to a Monte Carlo study, this proposal strongly overestimates the true parameter value when the two distributions overlap considerably.

The choice of the numerical value of the tolerance $\epsilon_I$ is based on a Monte Carlo simulation: we simulate $n=100$ observations from (\ref{eq:dens}) in the setup with $\xi=0.25$ and compute the MLEs with $\epsilon_I\in\{10^{-s}\}_{s=2,\dots,8}$. After repeating the experiment 100 times, we calculate the average computing time and the RMSE of all parameters (see Table \ref{tab:times}). The RMSEs suggest no significant improvement with $\epsilon_I<10^{-4}$, hence in the following we set $\epsilon_I=10^{-4}$. As a side remark, it is worth noting that the outcomes in Table \ref{tab:times} seem to denote a difficulty in identifying the Cauchy parameters, whose RMSEs remain quite large for all values of $\epsilon_I$.
\begin{table}[htbp]
	\centering
	\caption{MLE. Computing times (in seconds) and RMSE of the MLEs in 100 replications of the simulation experiment for threshold $\epsilon_I\in\{10^{-s}\}_{s=2,\dots,8}$. The true value of $\xi$ is 0.25 and the sample size is $n=100$.}
	\begin{tabular}{cccccccc}
		\hline 
		 $\epsilon_I$   &  Time   & $\mu_c$ & $\tau$ & $\mu$ & $\sigma$ & $\beta$ & $\xi$ \\ \hline
		 $10^{-2} $   &  5.26   &  5260.488  & 7771.369  & 0.260 &  0.235   & 1.902 &  0.271 \\
		 $10^{-3} $   &  11.43  &  4498.541  & 7423.595  & 0.215 &  0.241   & 1.777 &  0.255 \\
		 $10^{-4} $   &  17.44  &  3850.602  & 7006.747  & 0.182 &  0.187   & 1.334 &  0.232 \\
		 $10^{-5} $   &  52.09  &  4227.770  & 6312.171 & 0.173 &  0.201   & 1.389  &  0.242\\
		 $10^{-6} $   & 124.61  &  4489.436  & 7123.037 & 0.180 &  0.186   & 1.330  &  0.239\\ 
		 $10^{-7} $   & 233.14  &  3917.092  & 7107.936 & 0.176 &  0.192   & 1.332  &  0.241\\ 
		 $10^{-8} $   & 385.21  &  3945.839  & 7216.598 & 0.175 &  0.191   & 1.330  &  0.234\\ \hline
	\end{tabular}
	\label{tab:times}
\end{table}

We have implemented AMLE with and without summary statistics, as in sections \ref{sec:without} and \ref{sec:with}, respectively. In the former approach, the summary statistic is the empirical characteristic function, which has proved to be an effective choice in other cases (see, e.g., \citealp{bee18,bee22,beetaf22}). However, a small simulation experiment suggested a better performance of the full-data approach, both in statistical and computational terms (smaller root-mean-squared-error and shorter computing time). Hence, only the outcomes of the full-data approach based on the CvM distance are shown in this section.

The ABC sample size is $\ell=100$, with $k=5\cdot 10^5$, since a small pilot simulation with $k=2\cdot 10^6$ did not suggest any significant improvement in the statistical properties of the estimators with respect to $k=5\cdot 10^5$. The supports of the uniform priors are determined via the procedure outlined in Section \ref{sec:with}. Non-parametric bootstrap is replaced in this section by parametric bootstrap, since here we repeatedly simulate the true distribution.

In each experiment, we compute the bias, the standard deviation and the RMSE of all parameters, with $B=100$ replications. The AMLE outcomes shown in this section are based on the sample mean, since the results of the other approaches (``UK, ``MK'' and ``PUK'') are nearly identical.

Computer-intensive methods are usually characterized by a much larger computational burden, and AMLE makes no exception. When $n=500$, regardless of the true value of $\xi$, AMLE needs about 55 minutes, whereas MLE takes 32 seconds (see Table \ref{tab:times}). When $n=100$, these times reduce to 12.5 minutes for AMLE and 17 seconds for MLE\footnote{These computing times are based on \texttt{R} codes run on a Windows machine with an i7-6700 CPU \@ 3.40GHz. The simulation experiments have been run in parallel on the HPC computer cluster of the University of Trento, with a total time of about 22 hours for 100 replications when $n=500$.}. 

\subsection{Simulation results}

Tables \ref{tab:xi25} and \ref{tab:xi50} display bias, standard deviation and root-mean-squared-error (RMSE) of both AMLEs and MLEs of the parameters.
\begin{table}[htbp]
\centering
\caption{Case $\xi=0.25$. Bias, standard deviation and RMSE of the estimates obtained via AMLE and MLE in $B=100$ replications of the simulation experiment. The true value of $\xi$ is 0.25.}
\begin{tabular}{ccccccccc}
	\hline
	        $n$          &                       &      & $\mu_c$  & $\tau$ &  $\mu$   & $\sigma$ & $\beta$  &  $\xi$   \\ \hline
	\multirow{6}{*}{100} & \multirow{2}{*}{Bias} & AMLE & 0.554 & 0.879 & 0.090 & 0.114 & 0.375 & 0.008 \\
	                     &                       & MLE  & $-428.537$ & 686.732 & 0.053 & 0.034 & 0.300 & $-0.028$ \\
	                     &  \multirow{2}{*}{Sd}  & AMLE & 0.554 & 0.221 & 0.123 & 0.088 & 0.571 & 0.045  \\
	                     &                       & MLE  & 4273.488 & 6823.124 & 0.190 & 0.200 & 1.132 & 0.197  \\
	                     & \multirow{2}{*}{RMSE} & AMLE & 0.784 & 0.907 & 0.152 & 0.144 & 0.683 & 0.046   \\
	                     &                       & MLE  & 4294.921 & 6857.596 & 0.197 & 0.203 & 1.171 & 0.199 \\ \hline
	\multirow{6}{*}{500} & \multirow{2}{*}{Bias} & AMLE &  0.201 & 0.283 & 0.0179 &  0.035 &   0.184  & $-0.013$ \\
	                     &                       & MLE  & $-0.018$ & 0.429 & $-0.000$ & $-0.004$ & 0.029 & 0.007   \\
	                     &  \multirow{2}{*}{Sd}  & AMLE &  0.225 & 0.348 & 0.049 & 0.046 & 0.319 & 0.054   \\
	                     &                       & MLE  &  0.839 & 3.198 & 0.069 & 0.080 & 0.526 & 0.097   \\
	                     & \multirow{2}{*}{RMSE} & AMLE &  0.302 & 0.448 & 0.052 & 0.056 & 0.368 & 0.056   \\
	                     &                       & MLE  &  0.839 & 3.227 & 0.069 & 0.080 & 0.527 & 0.097   \\ 
	                     \hline
\end{tabular}
\label{tab:xi25}
\end{table}
\begin{table}[htbp]
	\centering
	\caption{Case $\xi=0.5$. Bias, standard deviation and RMSE of the estimates obtained via AMLE and MLE in $B=100$ replications of the simulation experiment. The true value of $\xi$ is 0.5.}
	\begin{tabular}{ccccccccc}
		\hline
		        $n$          &                       &      &  $\mu_c$   &  $\tau$  &  $\mu$   & $\sigma$ & $\beta$ &   $\xi$    \\ \hline
	\multirow{6}{*}{100} & \multirow{2}{*}{Bias} & AMLE &  0.881 & 2.084 & 0.049 & 0.087 & 0.780 & $-0.094$   \\
	                     &                       & MLE  &  0.677 & 0.447 & 0.018 & 0.019 & 0.652 & $-0.066$  \\
	                     &  \multirow{2}{*}{Sd}  & AMLE &  1.217 & 1.542 & 0.121 & 0.114 & 0.923 & 0.146 \\
	                     &                       & MLE  &  2.093 & 2.953 & 0.158 & 0.168 & 1.093 & 0.193    \\
	                     & \multirow{2}{*}{RMSE} & AMLE &  1.502 & 2.592 & 0.131 & 0.143 & 1.208 & 0.174 \\
	                     &                       & MLE  &  2.199 & 2.980 & 0.159 & 0.169 & 1.271 & 0.204 \\ \hline
	\multirow{6}{*}{500} & \multirow{2}{*}{Bias} & AMLE &  0.403 & 0.062 & 0.027 & 0.035 & 0.158 & $-0.005$ \\
	                     &                       & MLE  &  0.048 & 0.330 & 0.009 & 0.004 & 0.085 & $-0.001$ \\
	                     &  \multirow{2}{*}{Sd}  & AMLE &  0.443 & 0.605 & 0.059 & 0.053 & 0.413 & 0.075  \\
	                     &                       & MLE  &  0.708 & 1.251 & 0.067 & 0.063 & 0.496 & 0.083 \\
	                     & \multirow{2}{*}{RMSE} & AMLE &  0.599 & 0.608 & 0.065 & 0.062 & 0.442 & 0.075 \\
	                     &                       & MLE  &  0.709 & 1.293 & 0.068 & 0.063 & 0.503 & 0.083 \\ \hline
	\end{tabular}
	\label{tab:xi50}
\end{table}
The tables suggest that AMLE is always preferable to MLE in terms of RMSE, by a larger amount when $\xi=0.25$; in most cases, the gain in RMSE results from a smaller standard deviation of AMLE, which more than compensates a slightly larger bias. Moreover, the advantage is larger for small sample sizes.

For the case $\xi=0.25$, the simulated distributions of the AMLEs of the parameters are displayed in Figure \ref{fig:histo6c} (for $n=100$) and \ref{fig:histo6} (for $n=500$), along with the supports of the uniform distributions used in the AMLE algorithm; the corresponding plots when $\xi=0.5$ are very similar and therefore omitted. The histograms are approximately bell-shaped for both sample sizes, which explains why the four AMLE approaches yield very similar outcomes.
\begin{figure}[H]
	\begin{center}
		\includegraphics[width=.7\textwidth]{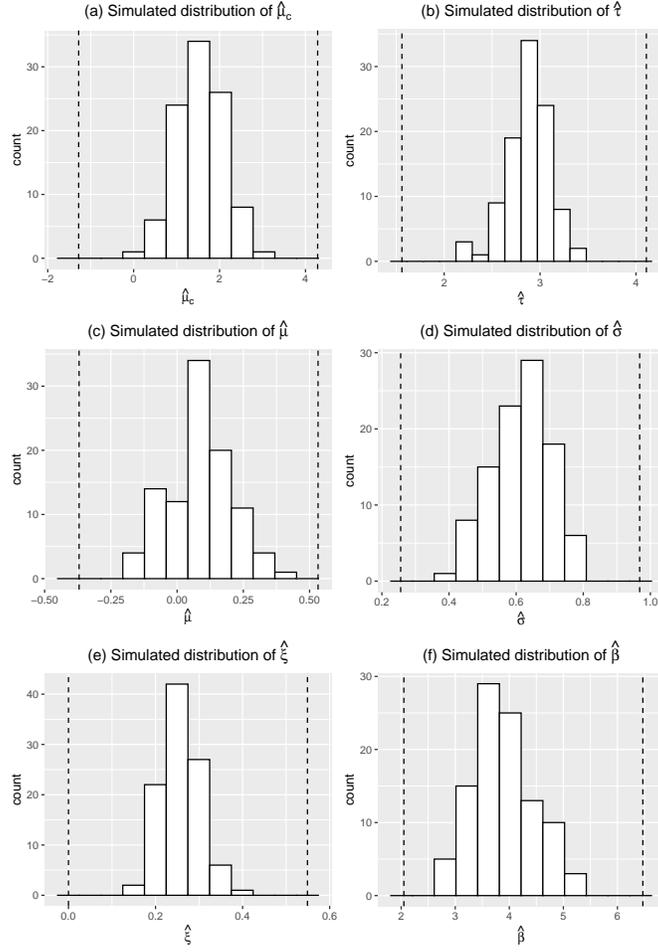}
		\caption{Simulated distributions of the AMLE estimators with $n=100$. The number of replications is $B=100$ and the true parameter values are $(\mu_c, \tau, \mu, \sigma, \xi, \beta) =(1,2, 0,0.5,0.25, 3.5)$. The dashed vertical lines are the lower and upper bounds of the uniform priors.}
		\label{fig:histo6c}
	\end{center}
\end{figure}
\begin{figure}[H]
	\begin{center}
		\includegraphics[width=.75\textwidth]{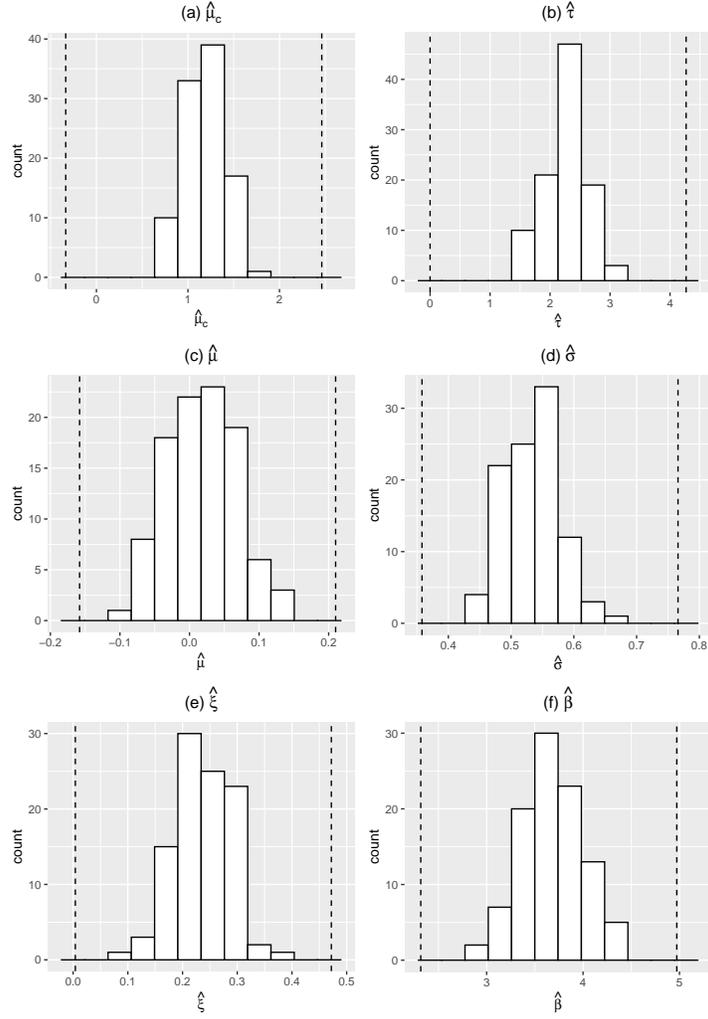}
		\caption{Simulated distributions of the AMLE estimators with $n=500$. The number of replications is $B=100$ and the true parameter values are $(\mu_c, \tau, \mu, \sigma, \xi, \beta) =(1,2, 0,0.5,0.25, 3.5)$. The dashed vertical lines are the lower and upper bounds of the uniform priors.}
		\label{fig:histo6}
	\end{center}
\end{figure}

Similarly to \cite{fri02}, our analysis suggests that estimation of the dynamic weight parameters is difficult, in particular with the MLE method: as can be seen from the large bias and standard deviation of $\hat{\mu}_c^{MLE}$ and $\hat{\tau}^{MLE}$ in Table \ref{tab:xi25}, there are cases where the numerical optimization of the log-likelihood function does not yield sensible outcomes. It is also worth noting that in a few cases (approximately 2-5\% of the replications, with the largest values when $\xi=0.25$ and $n=100$) the numerical maximization of the log-likelihood broke down without producing a result. In such cases we have discarded the sample and simulated the observations again.

\section{Empirical analysis}
\label{sec:emp}

\subsection{Operational risk}

An important application of the model considered in this paper is the analysis of loss distributions, since they are typically skewed and heavy-tailed. In many cases, the final goal is the estimation of risk measures such as the Value-at-Risk (VaR) or the Expected Shortfall (ES). For this purpose, the tail must be estimated with a high level of precision, so that accurate models of the tail are of paramount importance.

\begin{figure}[H]
	\begin{center}
		\includegraphics[width=.95\textwidth]{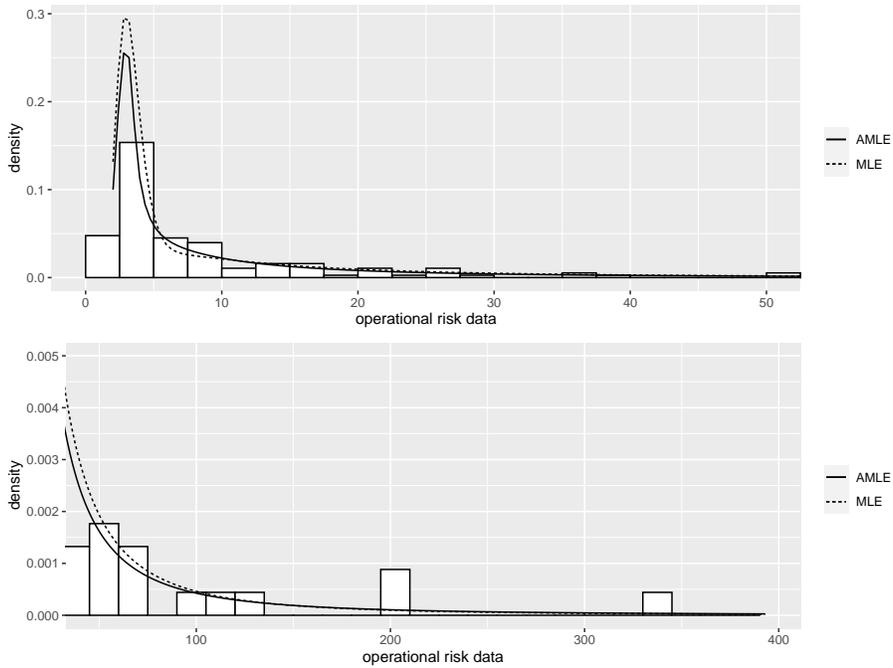}
		\caption{The operational risk data with superimposed the dynamic lognormal-GPD density estimated via AMLE (continuous) and MLE (dashed). The upper and lower panel respectively show the body ($x\le 50$) and the tail ($x> 50$). AMLE estimators are based on the ``PUK'' method.}
		\label{fig:ORdata}
	\end{center}
\end{figure}

In this section, we use operational risk losses collected at the Italian bank Unicredit between 2005 and 2014. In particular, we model the monetary amounts of the losses in the \emph{Business Disruption and System Failure} (BDSM) business class. The sample size is $n=152$. The data are displayed in Figure \ref{fig:ORdata}, along with the densities estimated via AMLE and MLE.
Table \ref{tab:OR_res} shows the MLEs and AMLEs; standard errors are computed via non-parametric bootstrap with $B=100$ replications. 

\begin{table}[htbp]
	\centering
	\caption{Parameter estimates and standard errors obtained via AMLE and MLE in the operational risk example. Standard errors are computed via non-parametric bootstrap with $B=100$ bootstrap replications.}
	\begin{tabular}{ccccccc}
		\hline
		&  $\mu_c$ & $\tau$ & $\mu$ & $\sigma$ & $\xi$ & $\beta$ \\
		\hline
		\multirow{2}{*}{AMLE M} & 3.882 & 0.587 & 1.309 & 0.330  & 0.921 & 5.557  \\
							  & (0.626) & (0.352) & (0.079) & (0.053) & (0.119) & (0.819) \\
		\hline
		\multirow{2}{*}{AMLE UK} &  3.410 & 0.435 & 1.303 & 0.330 &  0.912 & 5.466   \\
		& (12.742) & (0.673) & (0.082) & (0.055) & (0.137) &  (1.497)  \\
		\hline
		\multirow{2}{*}{AMLE MK} & 3.362 & 0.453 & 1.295 & 0.293 & 0.908 & 4.930 \\
		& (0.846) & (0.389) & (0.106) & (0.080) & (0.197) & (1.682)  \\
		\hline
		\multirow{2}{*}{AMLE PUK}   & 3.423 & 0.447 & 1.307 & 0.320 & 0.916 & 5.348 \\
		& (0.692) & (0.224) & (0.090) & (0.064)  & (0.160) & (1.503)  \\
		\hline
		\multirow{2}{*}{MLE} & 4.947 & 3.116 & 1.197 & 0.290 & 0.706 & 7.871 \\
							 & (1.402) & (1.276) & (0.103) & (0.082) & (0.191) & (2.836) \\
		\hline
	\end{tabular}
	\label{tab:OR_res}
\end{table}
There is a considerable difference between AMLE and MLE when estimating $\tau$, $\beta$ and, to a lesser extent, $\mu_c$. The large standard errors imply that, for these parameters, the precision is low, but in terms of variability AMLEs are definitely preferable. Notwithstanding the differences in the estimated values of some parameters, the densities in Figure \ref{fig:ORdata} are rather similar: this suggests that the errors in the estimates of individual parameters are likely to offset each other \citep{fri02}. Notice also that the observations above 50 in Figure \ref{fig:ORdata} are very likely to come from the GPD, since the expected values of the lognormal and GPD (computed using the PUK estimators) are as different as $\exp\{1.307+0.32^2/2\}=3.889$ and $5.348/(1-.916)=63.667$, respectively.

With regard to the four AMLE methods, there is some  difference mostly when estimating $\mu_c$ and $\tau$. To understand why, Figure \ref{fig:ORdistr} displays the distribution of the ABC samples corresponding to the six parameters.
\begin{figure}[H]
	\begin{center}
		\includegraphics[width=\textwidth]{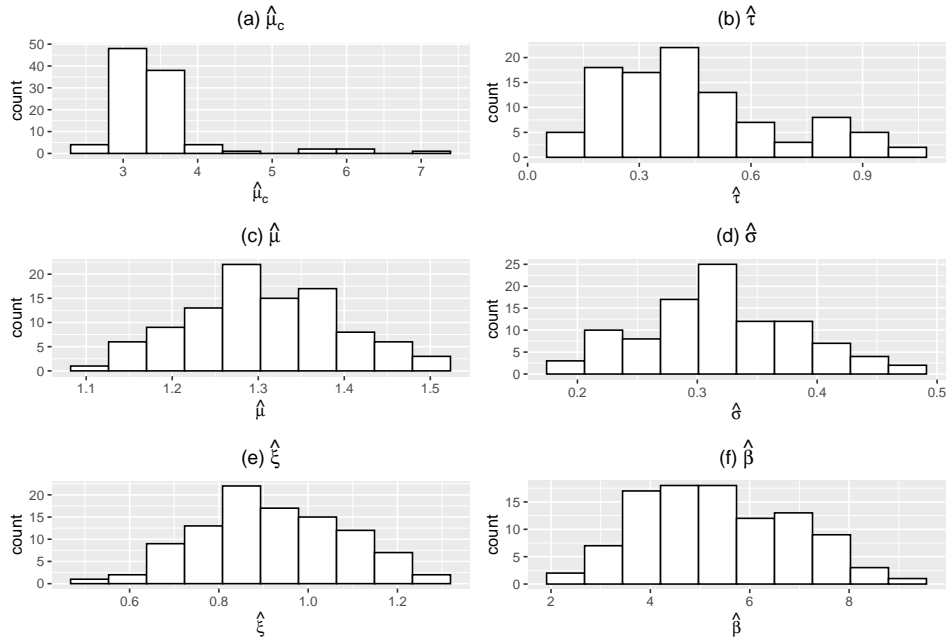}
		\caption{The ABC samples of the six parameters in the operational risk data.}
		\label{fig:ORdistr}
	\end{center}
\end{figure}
The distributions of the samples of $\hat{\mu}_c$ and $\hat{\tau}$ are the least symmetric. In such cases, it is not recommended to estimate the maximum with the sample mean, hence we suggest using either ``UK'' or ``PUK'', which are approximately identical. Notwithstanding the small sample size, ``MK'' also yields estimates close to ``UK'' and ``PUK''.

Tables \ref{tab:ORq} and \ref{tab:ORtail} respectively show quantiles at different levels and estimated tail probabilities $P(X\ge t)$ for various values of $t$. Note that, in this setup, the $\alpha$-quantile can be interpreted as the Value-at-Risk at level $\alpha$ (see, e.g., \citealp[Section 2.3.2]{mcn15}).

Given that we focus on the tail, as a benchmark we also compute the quantiles based on the GPD asymptotic approximation fitted to the excesses $x_i-u$, where $u$ is taken to be equal to the 90-th quantile, by means of the Peaks-over-Threshold (POT) method \citep[Section 5.3.2]{mcn15}. Since this is a theoretically well-grounded approach aimed at tail approximation, it is expected to be a reliable benchmark. Finally, we also report the sample quantiles of the observed data (Table \ref{tab:ORq}) and the empirical estimate of the tail probabilities $p_t=:\#\{x_i:x_i\ge t\}/n$ (Table \ref{tab:ORtail}).
\begin{table}[htbp]
	\centering
	\caption{Estimated quantiles obtained via AMLE, MLE and the POT method in the operational risk example. Empirical quantiles of the observed data are reported as well.}
	\begin{tabular}{cccccc}
		\hline
				&   50\% &   90\% &      95\% &       99\% &    99.5\% \\
		\hline
		{AMLE M} & 4.493 & 39.404 & 81.632 & 385.415 & 731.770 \\
		{AMLE UK} & 4.911 & 40.916 & 82.406 & 369.166 & 688.687 \\
		{AMLE MK} & 4.703 & 37.343 & 75.060 & 323.465 & 617.791 \\
		{AMLE PUK} & 4.907 & 40.298 & 82.216 & 390.626 & 750.067 \\
		{MLE} & 4.222 & 36.763 & 66.717 & 230.602 & 385.943 \\
		{GPD} & -     & 39.941 & 84.153 & 257.865 & 380.208 \\
		{EMP} & 4.990 & 37.176 & 73.569 & 274.323 & 355.255 \\
		\hline
	\end{tabular}
	\label{tab:ORq}
\end{table}
\begin{table}[htbp]
	\centering
	\caption{Estimated tail probabilities obtained via AMLE, MLE and the POT method in the operational risk example. Empirical tail probabilities of the observed data are reported as well.}
	\begin{tabular}{cccccccc}
		\hline
		         	&  100 & 150 & 200 & 300 & 400 & 500 & 600 \\
		\hline
		{AMLE M} & 0.038 & 0.025 & 0.019 & 0.012 & 0.009 & 0.007 & 0.005 \\
		{AMLE UK} & 0.041 & 0.027 & 0.020 & 0.013 & 0.009 & 0.007 & 0.006 \\
		{AMLE MK} & 0.037 & 0.025 & 0.018 & 0.012 & 0.009 & 0.007 & 0.006 \\
		{AMLE PUK} & 0.041 & 0.027 & 0.020 & 0.013 & 0.010 & 0.007 & 0.006 \\
		{MLE} & 0.030 & 0.018 & 0.012 & 0.007 & 0.005 & 0.003 & 0.003 \\
		{GPD} & 0.041 & 0.024 & 0.015 & 0.008 & 0.005 & 0.003 & 0.002 \\
		{EMP} & 0.039 & 0.026 & 0.026 & 0.013 & 0.000 & 0.000 & 0.000 \\
		\hline
	\end{tabular}
	\label{tab:ORtail}
\end{table}
Tables \ref{tab:ORq} and \ref{tab:ORtail} suggest that the AMLE approaches yield quite similar results; only MK gives slightly smaller estimated quantiles. For the smallest quantile levels (90 and 95\%) and the smallest thresholds ($\le 200$), the AMLE estimates are closer than MLE to both the GPD and the empirical estimates. On the other hand, when one moves farther into the tail, the MLE-based estimates are very close to the GPD-based estimates. It is interesting to notice that, even though the densities in Fig. \ref{fig:ORdata} look very close to each other, when the quantile level is high the difference between AMLE and MLE becomes non-negligible.

In terms of Expected Shortfall, Table \ref{tab:ORES} shows that the dynamic mixture, especially when estimated via AMLE, yields numerical values larger than both the empirical and the GPD-based measure. However, interpreting such a comparison in terms of ES with a small sample size requires some care, since ES is very sensitive to a few very large observations.

\begin{table}[htbp]
	\centering
	\caption{Estimated Expected Shortfall obtained via AMLE, MLE and the POT method in the operational risk example. The empirical ES of the observed data is reported as well.}
	\begin{tabular}{cccccc}
		\hline
				&   50\% &   90\% &      95\% &       99\% &    99.5\% \\
		\hline
		{AMLE M} & 80.632 &  346.086 &  634.897 & 2535.654 & 4555.332 \\
		{AMLE UK} & 85.142 &  368.395 &  680.077 & 2764.897 & 5010.873 \\
		{AMLE MK} & 80.161 & 343.524 &  628.964 & 2490.805 & 4438.181 \\
		{AMLE PUK} & 109.452 &  490.383 &  922.963 & 3961.587 & 7399.500 \\
		{MLE} & 38.756 &  142.389 &  235.664 &  716.620 & 1133.496 \\
		{GPD} & - & 137.637 & 216.876 & 528.209 & 747.477 \\
		{EMP} & 35.933 & 123.898 & 194.493 & 368.010 & 393.020 \\
		\hline
	\end{tabular}
	\label{tab:ORES}
\end{table}

Finally, to assess the quality of the VaR and ES estimates obtained, we use the Mean Absolute Relative Error (MARE) proposed by \cite{nad20}:
$$
MARE = \frac{1}{n_l}\sum_{i=1}^{n_l}\left|\frac{M_i-\hat{M}_i}{M_i}\right|,
$$
where $n_l$ is the number of levels and $M_i$ and $\hat{M}_i$ denote the empirical and estimated risk measure at the $i$-th level, respectively. Table \ref{tab:MAREOR} displays the results, which confirm that MLE is closer to the empirical values.

\begin{table}[htbp]
	\centering
	\caption{Mean absolute relative error for AMLE, MLE and the POT method in the operational risk example.}
	\begin{tabular}{ccccccc}
		\hline
		           &   M   &  UK   &  MK   &  PUK  &  MLE  &  GPD  \\ \hline
		{MARE VaR} & 0.264 & 0.210 & 0.223 & 0.219 & 0.103 & 0.087 \\
		{MARE ES}  & 7.316 & 6.262 & 8.248 & 9.298 & 3.001 & 1.670 \\ \hline
	\end{tabular}
	\label{tab:MAREOR}
\end{table}

\subsection{Metropolitan cities}

In the last two decades or so, city size data have often been the focus of a lively dispute between scholars: while some argue that the whole distribution is lognormal, others claim that the body is lognormal but the tail follows a Pareto-type distribution. See, e.g., \cite{dac19} and the references therein.

Here we study the distribution of the 2019 population estimate, divided by 10\,000, of the 415 US metropolitan areas computed by the US Census Bureau\footnote{www.census.gov/data/datasets/time-series/demo/popest/2010s-total-metro-and-micro-statistical-
areas.html.}; for an earlier investigation of US metropolitan areas, see \cite{gaib11}. The data and the two estimated densities are shown in Figure \ref{fig:metrodata}. Whereas in the previous application the two densities were very similar to each other (see Figure \ref{fig:ORdata}), in this case the MLE-based density gives less weight to the body and more weight to the tail; this fact is confirmed by the estimated quantiles in Table \ref{tab:metroq}.
\begin{figure}[H]
	\begin{center}
		\includegraphics[width=\textwidth]{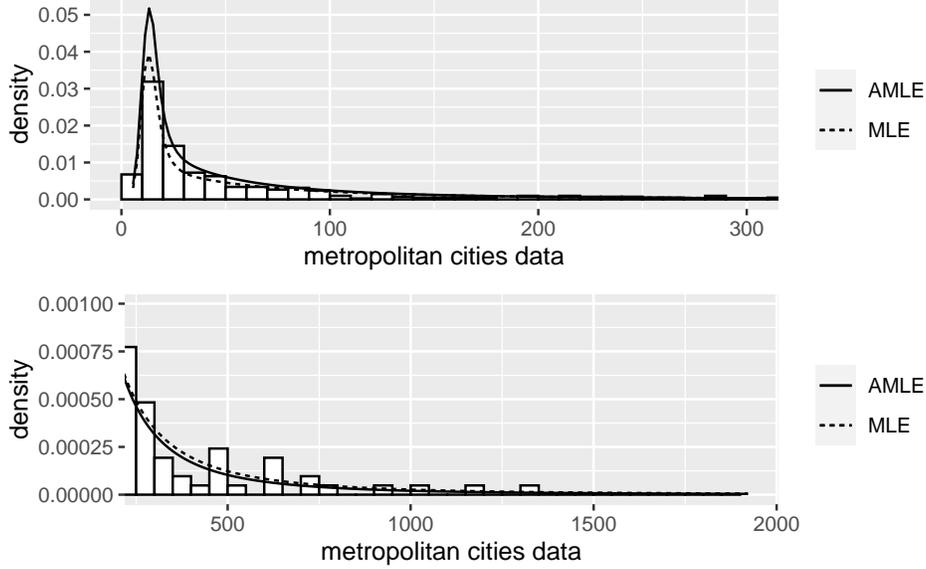}
		\caption{The metropolitan cities data with superimposed the dynamic lognormal-GPD density estimated via AMLE (continuous) and MLE (dashed). The upper and lower panel respectively show the body ($x\le 300$) and the tail ($x> 300$). AMLE estimators are based on the ``PUK'' method.}
		\label{fig:metrodata}
	\end{center}
\end{figure}

\begin{table}[htbp]
	\centering
	\caption{Parameter estimates and standard errors obtained via AMLE and MLE in the metropolitan cities example. Standard errors are computed via non-parametric bootstrap with $B=100$ bootstrap replications.}
	\begin{tabular}{ccccccc}
		\hline
		&  $\mu_c$ & $\tau$ & $\mu$ & $\sigma$ & $\xi$ & $\beta$ \\
		\hline
		\multirow{2}{*}{AMLE M} & 26.493 & 2.667 & 2.869 &  0.455 & 0.595 & 55.770 \\
	   							   & (0.913) & (0.288) & (0.039) & (0.021) & (0.078) & (2.427) \\
		\hline
		\multirow{2}{*}{AMLE UK} &  24.361 & 3.991 & 2.873 & 0.456 & 0.590 & 55.216   \\
									& (1.367) & (0.391) & (0.039) & (0.022) & (0.081) & (2.354) \\
		\hline
		\multirow{2}{*}{AMLE MK} & 23.975 & 2.499 & 2.887 & 0.475 & 0.583 & 50.971 \\
									& (1.775) & (0.685) & (0.049) & (0.037) & (0.107) & (3.359) \\
		\hline
		\multirow{2}{*}{AMLE PUK}   & 24.714 & 3.246 & 2.895 & 0.467 & 0.547 & 54.523 \\
									& (1.470) & (0.501) & (0.047) & (0.038) & (0.098) & (2.526) \\
		\hline
		\multirow{2}{*}{MLE} & 15.002 &  4.504 & 2.829  & 0.359 & 0.680 & 56.602 \\
							& (2.992) & (0.426) & (0.079) & (0.046) & (0.118) & (9.570) \\
		\hline
	\end{tabular}
	\label{tab:metro_res}
\end{table}

Parameter estimates and standard errors are shown in Table \ref{tab:metro_res}. Once again, the Cauchy parameters are difficult to estimate, and one can notice large differences between AMLE and MLE in the estimated value of the location parameter $\mu_c$. 
In terms of variability, the standard errors of AMLEs are almost always smaller than those of MLEs, and for some parameters the gain is considerable.

Analogously to the previous section, Table \ref{tab:metroq} shows selected quantiles estimated via both methods, and Table \ref{tab:metrotail} reports estimated tail probabilities $P(X\ge t)$ for various values of $t$.
\begin{table}[htbp]
	\centering
	\caption{Estimated quantiles obtained via AMLE, MLE and the POT method in the metropolitan cities example. Empirical quantiles of the observed data are reported as well.}
	\begin{tabular}{cccccc}
		\hline
		&   50\% &   90\% &      95\% &       99\% &    99.5\% \\
		\hline
		{AMLE M} & 29.417 & 205.714 & 359.239 & 1108.211 & 1747.710 \\
		{AMLE UK} & 27.615 & 190.465 & 326.674 & 1002.483 & 1625.888 \\
		{AMLE MK} & 28.791 & 204.030 & 347.917 & 1090.096 & 1708.617 \\
		{AMLE PUK} & 28.302 & 195.501 & 337.762 & 1069.634 & 1716.508 \\
		{MLE} & 34.223 & 273.854 & 487.290 & 1567.529 & 2576.189 \\
		{GPD} & -     & 227.171 & 366.553 &  884.398 & 1231.582 \\
		{EMP} & 27.32 & 225.09 & 342.88 &  919.46 & 1170.91 \\
		\hline
	\end{tabular}
	\label{tab:metroq}
\end{table}

\begin{table}[htbp]
	\centering
	\caption{Estimated tail probabilities obtained via AMLE, MLE and the POT method in the metropolitan cities example. Empirical tail probabilities of the observed data are reported as well.}
	\begin{tabular}{ccccccccc}
		\hline
				&   200 & 400 & 600 & 700 & 800 & 900 & 1000 &  1500 \\
		\hline
		{AMLE M} & 0.101 & 0.041 & 0.024 & 0.019 & 0.0153 & 0.0129 & 0.011 & 0.006 \\
		{AMLE UK} & 0.095 & 0.039 & 0.022 & 0.018 & 0.015 & 0.012 & 0.011 & 0.006 \\
		{AMLE MK} & 0.099 & 0.041 & 0.023 & 0.019 & 0.0153 & 0.0126 & 0.011 & 0.006 \\
		{AMLE PUK} & 0.097 & 0.041 & 0.023 & 0.019 & 0.016 & 0.014 & 0.012 & 0.007 \\
		{MLE} & 0.138 & 0.062 & 0.037 & 0.030 & 0.026 & 0.022 & 0.019 & 0.011 \\
		{GPD} & - & 0.043 & 0.021 & 0.016 & 0.012 & 0.001 & 0.008 & 0.003 \\
		{EMP} & 0.123 & 0.046 & 0.029 & 0.019 & 0.012 & 0.012 & 0.001 & 0.002 \\
		\hline
	\end{tabular}
	\label{tab:metrotail}
\end{table}

Tables \ref{tab:metroq} and \ref{tab:metrotail} suggest that the AMLE-based estimated quantiles and tail probabilities are closer than the MLE-based quantities to both the GPD-based and the empirical measures. MLE-based quantiles and tail probabilities are considerably larger, especially at high levels and thresholds. The ES measures in Table \ref{tab:MetroES} essentially confirm these remarks.

\begin{table}[htbp]
	\centering
	\caption{Estimated Expected Shortfall obtained via AMLE, MLE and the POT method in the metropolitan cities example. The empirical ES of the observed data is reported as well.}
	\begin{tabular}{cccccc}
		\hline
				&   50\% &   90\% &      95\% &       99\% &    99.5\% \\
		\hline
		{AMLE M} & 194.015 &  648.499 & 1036.208 & 2944.561 & 4530.217 \\
		{AMLE UK} & 199.711 &  680.113 & 1097.263 & 3222.347 & 5102.488 \\
		{AMLE MK} & 190.648 &  631.351 &  996.834 & 2693.999 & 3989.616 \\
		{AMLE PUK} & 201.276 &  685.638 & 1108.676 & 3272.185 & 5207.477 \\
		{MLE} & 303.493 &  1106.343 &  1857.966 &  6094.621 & 10193.591 \\
		{GPD} & - & 516.622 &  747.522 & 1605.441 & 2180.643 \\
		{EMP} & 162.353 &  495.473 &  720.431 & 1275.270 & 1475.528 \\
		\hline
	\end{tabular}
	\label{tab:MetroES}
\end{table}

Table \ref{tab:MAREmetro} displays the MARE results, which allow us to conclude that AMLE-based risk measures are preferable to MLE-based measures.

\begin{table}[htbp]
	\centering
	\caption{Mean absolute relative error for AMLE, MLE and the POT method in the metropolitan cities example.}
	\begin{tabular}{ccccccc}
		\hline
		           &   M   &  UK   &  MK   &  PUK  &  MLE  &  GPD  \\ \hline
		{MARE VaR} & 0.175 & 0.174 & 0.163 & 0.174 & 0.611 & 0.042 \\
		{MARE ES}  & 2.790 & 2.892 & 2.937 & 3.331 & 5.381 & 1.021 \\ \hline
	\end{tabular}
	\label{tab:MAREmetro}
\end{table}

In this application it is relevant to study the degree of overlap of the two component distributions, and the possible presence of a GPD tail. Most of the investigations carried out in the literature try to find a threshold where the power-law distribution starts, as if the two distributions did not overlap. This way of proceeding is not correct, since the two distributions actually overlap. The present approach, however, does not suffer from this drawback.

\cite{bee22a} analyzes the same dataset; even though the model is different, it is interesting to compare the number of Pareto observations found in that paper to the number of observations that are likely to be GPD in the present dynamic mixture model.

The likelihood of an observation being GPD can be measured by means of the estimated weight $p(x;\mu_c,\tau)$, by setting a high probability level $\alpha$ and finding the smallest $x:p(x;\mu_c,\tau)>\alpha$. This value ($x_{\alpha}$, say) can be interpreted as a threshold above which almost all the observations are GPD. Figure \ref{fig:metroC} shows the weights estimated via AMLE ``PUK'' and MLE.
\begin{figure}[!h]
	\begin{center}
		\includegraphics[width=\textwidth]{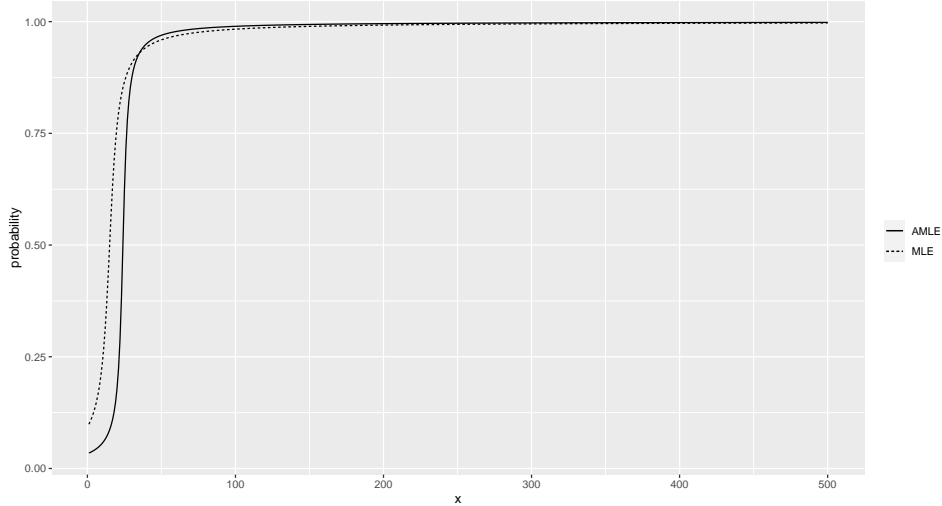}
		\caption{Estimated dynamic weight functions obtained via AMLE ``PUK'' and MLE in the metropolitan cities example.}
		\label{fig:metroC}
	\end{center}
\end{figure}

Table \ref{tab:metrodist} displays thresholds $x_\alpha:p(x_\alpha;\mu_c,\tau)=\alpha$ estimated with all the methods for various values of $\alpha$. Given $\alpha$, the probability that an observation $x>x_\alpha$ is generated from the GPD is thus larger than $\alpha$. The number of observations exceeding a given $x_\alpha$ is reported in parentheses: for example, there are 304 observations larger than 15.26. In each row, these numbers give an idea of the number of GPD observations, since in the two rightmost columns only 1\% and 0.5\% of the observations are not GDP. For comparison, by fitting a lognormal-Pareto model with fixed mixing weight to the same dataset, \cite{bee22a} found 214 Pareto observations.

For high values of $\alpha$, the difference between the estimation methods is non-negligible, but in all cases there is a rather large number of GPD observations, consistently with the findings in \cite{bee22a}.
\begin{table}[htbp]
	\centering
	\caption{Estimated values $x:p(x;\mu_c,\tau)>\alpha$ obtained with all the methods for various values of $\alpha$. The numbers of observations exceeding the corresponding value $x_\alpha$ are reported in parentheses.}
	\begin{tabular}{ccccc}
		\hline
		           &  0.90  &  0.95  & 0.99  & 0.995   \\ \hline
		 {AMLE M}  & 10.313 & 11.129 & 15.260 (304) & 25.883 (219)\\
		{AMLE UK}  & 10.401 & 11.577 & 18.167 (263) & 52.672 (137) \\
		{AMLE MK}  & 10.227 & 10.983 & 14.966 (311) & 22.925 (232) \\
		{AMLE PUK} & 10.313 & 11.351 & 16.753 (288) & 32.620 (184) \\
		  {MLE}    & 10.081 & 11.129 & 19.769 (256) & 68.386 (113) \\
		 \hline
	\end{tabular}
	\label{tab:metrodist}
\end{table}

\section{Conclusion}
\label{sec:concl}

In this paper we have proposed a simulation-based procedure for the estimation of the parameters of a dynamic mixture. The motivation of this choice is that numerical MLE of this model is complicated by the need of approximating numerically the normalizing constant of the distribution.

To this aim, we have implemented an approximate maximum likelihood approach in the lognormal-GPD case, using the Cramér-von Mises distance to measure the discrepancy between observed and simulated samples. On the theoretical side, we have proved pointwise convergence of the AMLE approximation to the likelihood. In addition, we have given conditions under which the mode of the approximation converges to the mode of the likelihood. Simulation results suggest that AMLE outperforms MLE in terms of RMSE, albeit with a higher computational cost. Two empirical applications confirm that the approach is successful at fitting skewed datasets in economics and finance.

Various issues need further research. First, it is possible to carry out a thorough comparison between the dynamic mixture considered in this paper and the lognormal-Pareto model mentioned in Section \ref{sec:model} (\citealp{scoll07}), in terms of both goodness of fit and numerical difficulty of the estimation procedures. 
Second, since the main issue with standard MLE is the evaluation of the normalizing constant, this approach could be dramatically improved by finding a weighting function that allows one to solve the integral in closed form, possibly at the price of some loss in flexibility. Third, computing bootstrap standard errors is time-consuming. \cite{wan16} have developed the score vector and Hessian matrix for multivariate \textit{t} mixtures, extending earlier work by \cite{bol09} for multivariate Gaussian mixtures. Since their results cannot be directly applied to our setup, which is somewhat different, further research on a possible extension to dynamic mixtures is necessary.
Fourth, setting up a more parsimonious model may help to reduce estimation difficulties: one possibility would be to replace the Cauchy cdf with a single-parameter cdf, such as the exponential. Similarly, exploring the performance of a model based on distributions different from the lognormal and the GPD may be of interest.
Finally, when the scale parameter of the Cauchy cdf tends to zero, the dynamic weight converges to the heavyside function: it may therefore be of interest to exploit the estimated value of this parameter to select the type of weight that is more appropriate for a given dataset.

\bigskip
\noindent\textbf{Acknwoledgements} The \texttt{R} package \texttt{LNPar}, containing the data used in the paper and the implementation of the methodology, is available at the address https://github.com/marco-bee/LNPar.

\end{document}